\documentclass[11pt,a4paper]{article}
\pdfoutput=1
\usepackage{amsmath,amssymb}
\usepackage{epsfig,graphicx}
\usepackage{subfigure}
\usepackage{graphicx}
\usepackage{rotating}
\usepackage{cancel}
\usepackage{bm}
\usepackage{color}
\usepackage{comment}
\usepackage{cite}

\def\beq{\begin{equation}}
\def\eeq{\end{equation}}

\begin{document}
\numberwithin{equation}{section}
\title{{\normalsize  IPPP/12/37 ~ DCPT/12/74\hfill\mbox{}\hfill\mbox{}}\\
\vspace{2.5cm} \LARGE{\textbf{The TeV Dawn of SUSY Models\\ -- \\Consequences  for Flavour and CP \\[1cm]}}}
\author{\Large
Joerg Jaeckel\footnote{{\bf e-mail}: joerg.jaeckel@durham.ac.uk},
\
and Valentin V. Khoze\footnote{{\bf e-mail}: valya.khoze@durham.ac.uk}
\\[2ex]
\small{\em Institute for Particle Physics Phenomenology,
Department of Physics}
\\
\small{\em Durham University, Durham DH1 3LE, UK}\\[1.5ex]
}
\date{}
\maketitle

\begin{abstract}
\noindent Direct searches and the hints for the Higgs at $\sim 125\,{\rm GeV}$ put increasing pressure
on simple models of SUSY breaking, in particular (but not exclusively) on
those that automatically solve the flavour problem, such as gauge mediation.
SUSY-breaking parameters are pushed to higher and higher values, increasing the fine-tuning required to achieve
electroweak symmetry breaking at the observed scale. In this note we consider the situation for models which do not attempt to
solve the flavour problem. To treat them on equal footing, we consider the combined fine-tuning arising from electroweak symmetry breaking
as well as
from fulfilling flavour constraints. We also consider CP. We find that for an anarchic flavour (and CP) structure of the soft SUSY breaking terms, the minimum in the fine-tuning occurs at a scale of a few TeV. This is consistent with the current experimental situation and
leads to motivating conclusions for direct searches as well as future flavour and CP measurements.
\end{abstract}

\newpage

\section{Introduction}

Supersymmetric extensions of the Standard Model face several
challenges arising from experimental constraints. In particular, SUSY models have to reproduce:
\begin{itemize}
\item[(i)]{} the measured scale of electroweak symmetry breaking, characterised by the mass of the Z boson
\item[(ii)]{} the absence of additional (to the Standard Model) unsuppressed flavour-changing effects
\item[(iii)]{} experimental constraints on masses of superpartners
\item[(iv)]{} constraints arising from the Higgs boson
\item[(v)]{} the absence of additional unsuppressed CP violation.
\end{itemize}
(i) is explained naturally when the superpartners of Standard Model particles have masses of the order of the electroweak scale whereas (ii)
can be explained if the SUSY-breaking interactions are flavour universal as, e.g., in models of gauge mediation.
Models of gravity mediation, however, generically
suffer\footnote{Flavour universality in some gravity mediated models can be achieved by first assuming a minimal form of
the Kahler potential which allows one to write down flavour universal soft terms at $M_{\rm Pl}$. Then a symmetry can be imposed which would prevent
detuning of this  flavour universal structure during RG evolution from $M_{\rm Pl}$ down to the flavour symmetry freezing scale $M_{\rm fl}$.
In \emph{generic} gravity mediation we do not impose the Minimal Flavour Violation hypothesis \cite{D'Ambrosio:2002ex} to hold at $M_{\rm fl}$.}
 from flavour changing neutral currents and require a fine-tuned
choice of parameters to fulfill (ii).

Over the last two years LHC has severely tightened the constraints
on the SUSY partners requiring many of their masses to be
significantly above the electroweak scale. For example searches
for jets and missing energy~\cite{jetsmet} require that typical
light squark and gluino masses are of the order of a TeV or
beyond. Moreover, the Higgs around $125\,{\rm GeV}$ also
pushes SUSY breaking parameters towards larger values (assuming that the $125\,{\rm GeV}$ resonance observed at the LHC \cite{HiggsCERN} is an elementary Higgs scalar).
Accordingly, an explanation of the electroweak scale requires significant
cancellations between larger parameters indicating a fine-tuned
situation. In models of gauge mediation (see e.g.
\cite{Giudice:1998bp} and references therein) this problem is
especially severe (for recent discussion see
Refs.~\cite{Grellscheid:2011ij,Draper:2011aa,talk1,feng}).

Indeed in the simplest models of gauge mediation, a Higgs mass of $\sim 125\,{\rm GeV}$ requires squark masses at or above
$\sim 10\,{\rm TeV}$, resulting in a very high level of fine-tuning (see also below).
Models of gravity mediation allow for $A$-terms at the high scale that ultimately can help with the Higgs mass. These
typically allow for somewhat lower values of the
superpartner masses, but still in the TeV range to conform to current jets and missing energy constraints~\cite{jetsmet}
for gluinos and squarks(at least for the first two generations, but in many simple models the third generation masses
are not much below those).

\medskip

In light of this TeV dawn, it is timely to re-examine
the flavour and the CP violation issues  and quantify the associated fine-tuning together with the one
arising from electroweak symmetry breaking. This is what we would like to achieve in  this note.\footnote{Our results were first
outlined in the talk \cite{talk1}. In the same spirit a similar approach has also been used in the very recent paper~\cite{feng}
to avoid potential CP issues in gauge mediation. A high SUSY breaking scale in gauge mediation may also be conducive to addressing dark matter~\cite{feng,okada}.}
For models with an a priori anarchic flavour structure,
flavour constraints require a large amount of fine-tuning when the superpartner masses are low, but when the masses are higher,
these constraints are ameliorated due to decoupling. A similar behavior can be found for constraints on CP
violation.
On the other hand, electroweak symmetry breaking prefers small values of the superpartner masses.
We find that the combined fine-tuning is minimal in an intermediate region around a few TeV.
Theoretically this suggests that the conceptually simplest models of gravity mediation without flavour/CP structure should be re-considered.
Experimentally, the preference for TeV-scale masses still preserves a hope of finding observable effects in direct searches of supersymmetry,
but importantly also in flavour and CP violation measurements.

\medskip

Since both flavour and electroweak symmetry breaking are most directly influenced by the sfermions, the preference for
 the TeV scale is most direct for their masses. For this reason, in this paper we will predominantly concentrate on sfermion masses.
 What about gaugino masses? There are two options.
 Without any specific model building or symmetry assumptions one would expect that gauginos
 have masses of the same order as sfermions and hence our sfermion-mass-based analysis below is unaffected. An alternative theoretical
 option is to arrange for the gauginos
 to be protected by a symmetry, e.g. R-symmetry, which allows them to have naturally smaller masses than scalars.
 In this case one could perform a similar additional
 analysis for observables directly influenced by the gaugino masses.
 For completeness we will briefly discuss two gaugino-sensitive observables in the last section before the conclusions.

While the detailed analysis of the combined fine-tuning as a function of few (or all) of the
 soft SUSY parameters can be contemplated and is in principle achievable, it would go beyond the scope of this paper which aims to provide
 a simple analysis of combined fine-tunning
 effects in generic uncomplicated SUSY models. In particular, we do not consider neither the so-called natural SUSY models with
 $m_{\rm stop} \ll m_{\rm squark}$ nor the split SUSY models
with $m_{\rm gaugino} \ll m_{\rm squark}$ so that we can retain a degree of simplicity and model-independence in our discussion.

\section{The flavour problem and fine-tuning}\label{sect:fl-ew}

One of the strongest constraints on SUSY model building arises from the exclusion of flavour changing neutral currents.
In the following we will consider two such constraints. This is by no means a complete set but rather it should be taken as an example
of the type of effects that can be expected when we include flavour in the discussion.

A simple but very strong constraint arises from the flavour changing process
$\mu\to e\,\gamma$. The branching ratio for this process given by (see, e.g.,~\cite{primer})
\begin{equation}
\label{muegamma}
{\rm BR}(\mu\to e\,\gamma)\approx 10^{-6} C \left(\frac{|m^{2}_{\tilde{\mu}_{R}\tilde{e}_{R}}|}{m^{2}_{\tilde{l}_{R}}}
\right)^2\left(\frac{100\,{\rm GeV}}{m_{\tilde{l}_{R}}}\right)^4< 2.4 \times10^{-12},
\end{equation}
and is very tightly constrained~\cite{Adam:2011ch} as can be seen on the right hand side.
Here $C$ is a function of the bino to slepton mass ratio and varies between $15$ and $0.1$ for bino masses much smaller
than the slepton mass to $m_{\tilde{B}}=2m_{\tilde{L}_{R}}$. The right hand side gives the current experimental constraint.

From Eq.~\eqref{muegamma} we can see that for $m_{\tilde{l}_{R}}\sim 100\,{\rm GeV}$ the flavour-changing off-diagonal mass term,
$m^{2}_{\tilde{\mu}_{R}\tilde{e}_{R}},$ is constrained to be much smaller than the diagonal mass term. Indeed one needs
$m^{2}_{\tilde{\mu}_{R}\tilde{e}_{R}}/m^{2}_{\tilde{l}_{R}}\lesssim 10^{-3}$.

In models that have no intrinsic mechanism to solve the flavour problem, this corresponds to a significant fine-tuning.
Note, however, that the constraints become weaker as the overall mass scale of the superpartners is increased
and the superpartners decouple.
Let us try to quantify this.

A standard measure of fine-tuning for a given observed, or limited from above, quantity $\Delta$, which depends on a parameter $a$, is given
by~\cite{barbieri,strumia},
\begin{equation}
\label{barbieri}
F^{a}_{\Delta}=\left|\frac{\partial\log(\Delta(a))}{\partial\log(a)}\right|=\left|\frac{a}{\Delta(a)} \frac{\partial \Delta(a)}{\partial a}\right|.
\end{equation}

\begin{figure}[t!]
\begin{center}
\includegraphics[width=0.47\textwidth]{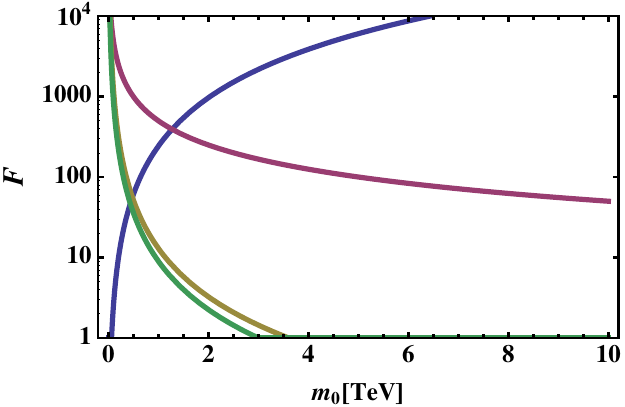}
\end{center}
\caption{Fine-tuning arising from different observables and parameters as a function of the overall soft SUSY breaking
scalar scale $m_{0}$. %(for details see Section~\ref{sect:combine}).
The blue curve corresponds to the electroweak fine-tuning in $m^{2}_{H_{u}}$. The yellow line is for the flavour changing
decay $\mu\to e\,\gamma$ with respect to the off-diagonal slepton mass term $m^{2}_{\tilde{\mu}_{R}\tilde{e}_{R}}$, red gives
the fine-tuning for $K^{0}\leftrightarrow\bar{K}^{0}$ oscillations with respect to the off-diagonal squark mass $m^{2}_{\tilde{s}\tilde{d}}$,
and finally the green curve arises from the CP violation fine-tuning discussed in Section~\ref{sect:CP}.}
\label{fig:finetuningindi}
\end{figure}

For example, the condition of electroweak symmetry breaking in the MSSM gives the well-known expression for the mass of the Z-boson in the limit of large $\tan(\beta)$,
\begin{equation}
m^{2}_{Z}\approx -2(m^{2}_{H_{u}}+|\mu|^2).
\end{equation}
We now view $m^{2}_{H_{u}}$ as the free parameter, and for the measure of fine-tuning of $m^{2}_{Z}$ we have
\begin{equation}
F^{m^{2}_{H_{u}}}_{m^{2}_{Z}}=2\frac{|m^{2}_{H_{u}}|}{m^{2}_{Z}}.
\label{eq:2.4}
\end{equation}
Whenever the Higgs soft mass parameter $m^{2}_{H_{u}}$ is much bigger than $m^{2}_{Z}$ we get a considerable degree of fine-tuning.
Taking, e.g., the soft-masses of the order of $10\,{\rm TeV}$, as
suggested \cite{Grellscheid:2011ij,talk1} in
vanilla models of gauge mediation, such as e.g. pure GGM models \cite{pGGM},
we get a fine-tuning of the order of $\sim10000$. This can also be seen from the blue line in Fig.~\ref{fig:finetuningindi} where we plot the fine-tuning as a function of the scalar mass scale.
The same (high) value is obtained when considering the variation with respect to $|\mu^2|$.

Note, however, that we could re-parameterise in terms of $x=m^{2}_{H_{u}}+|\mu|^2$ and $y=m^{2}_{H_{u}}-|\mu|^2$. In terms of these variables
the fine-tuning as measured with Eq.~\eqref{barbieri} is $F^{x}_{m^{2}_{Z}}=1$ and $F^{y}_{m^{2}_{Z}}=0,$ independent of the actual values
of $m^{2}_{H_{u}}$ and $|\mu^2|$.
In other words, the measured fine-tuning strongly depends on the choice of parameterisation\footnote{The smallness of the fine-tuning with this choice of parameterisation
is also due to the fact that there is no scale to which the size of $x=m^{2}_{H_{u}}-|\mu|^2$ is compared.
There is no reference made to the absolute scale that is expected for this variable, i.e. it is for example much smaller than the Planck scale.
This is similar to the issue discussed below for the smallness of flavour changing terms.}.
Indeed, this is as it should be because a theory that explains the hierarchy between the scale $m^2_{Z}$ and the scales
$m^2_{H_{u}}$ and $|\mu^2|$, will provide
a parameterisation that relates $m^{2}_{H_{u}}$ and $|\mu^2|$.
Therefore fine-tuning depends on the choice of what one considers fundamental parameters.
Indeed it can already make a difference at which scale one chooses the soft SUSY-breaking parameters because the RG evolution
provides a re-parameterisation in terms of the original high scale parameters. An example where this is relevant is the so-called hyperbolic branch of SUGRA models~\cite{focus}.
For example in the focus point\footnote{Recent LHC data~\cite{LHCnew}
severely depletes the focus point region (compare~\cite{focus} with~\cite{LHCnew}).}
and focus surface regions the relations between the high-scale parameters lead to an RG evolution such that the overall fine-tuning is significantly reduced
to the one expected from a naive consideration at the electroweak scale.

Let us now apply this reasoning to the flavour issue.
Naively we could consider the fine-tuning of the branching ratio given in Eq.~\eqref{muegamma} with respect to the off diagonal elements.
It is straightforward to see that this returns a small value of the fine-tuning independent of the tightness of the constraint on the branching ratio
and the corresponding very small value for the ratio between the diagonal and off-diagonal mass terms,
$m^{2}_{\tilde{\mu}_{R}\tilde{e}_{R}}/m^{2}_{\tilde{l}_{R}}$. This is a consequence of the fact that this parameterisation makes explicit
an especially symmetric situation on the level of the theory at the electroweak scale, namely an alignment between the flavour structure
in the fermions and that in the sfermions. However,
this alignment is not naturally expected in models of gravity mediation (see e.g. \cite{primer}) and can also be destroyed by the RG
flow~\cite{Calibbi:2012yj}.
Without additional structure it is therefore to be considered as accidental.
Indeed in absence of a symmetry a small parameter like in the above situation $m^{2}_{\tilde{\mu}_{R}\tilde{e}_{R}}/m^{2}_{\tilde{l}_{R}}\lesssim 10^{-3}$
suggests a corresponding fine-tuning.

A simple remedy to account for this fine-tuning is to shift the unnaturally small parameter by the expected size of this parameter.
In the above branching ratio we can, e.g. consider the new variable $x^2_{\tilde{\mu}_{R}\tilde{e}_{R}}=m^{2}_{\tilde{\mu}_{R}\tilde{e}_{R}}+m^{2}_{\tilde{l}_{R}}$.
This linear combination accounts for the fact that the diagonal and off-diagonal terms are expected to be of a similar order of magnitude
and can be played against each other in fine-tuning.
In terms of this  parameter $x^2_{\tilde{\mu}_{R}\tilde{e}_{R}}$ we find that the fine-tuning measure is,
\begin{equation}
\label{fine1}
F^{x^{2}_{\tilde{\mu}_{R}\tilde{e}_{R}}}_{BR(\mu\to e\,\gamma)}=2 \frac{m^{2}_{\tilde{l}_{R}}}{m^{2}_{\tilde{\mu}_{R}\tilde{e}_{R}}}
\end{equation}
which grows large as the ratio between diagonal and off diagonal terms decreases to unnaturally small values.

Solving Eq.~\eqref{muegamma} to obtain a constraint on the off-diagonal terms and inserting into \eqref{fine1} we get for the finetuning,
\begin{equation}
F^{x^{2}_{\tilde{\mu}_{R}\tilde{e}_{R}}}_{BR(\mu\to e\,\gamma)}\approx 13\,\,C^{1/2}\left(\frac{2.4\times10^{-12}}{BR(\mu\to e\,\gamma)}\right)^{1/2}
\left(\frac{1000\,{\rm GeV}}{m_{\tilde{l}_{R}}}\right)^2.
\label{eq:2.6}
\end{equation}
In contrast to the fine-tuning in the Z-boson mass this fine-tuning {\emph decreases} as the superpartner masses increase.
This can also be seen from the yellow line in Fig.~\ref{fig:finetuningindi}.

Another strong constraint involving strongly interacting sparticles arises from the mixing of neutral $K$ mesons, $K^{0}\leftrightarrow\bar{K}^{0}$,
Numerically~\cite{Ciuchini:1998ix,primer} this gives approximately
\begin{equation}
\frac{|{\rm Re}[m^{2}_{\tilde{s}_{R}\tilde{d}_{R}}m^{2}_{\tilde{s}_{L}\tilde{d}_{L}}]|^{1/2}}{m^{2}_{\tilde{q}}}\left(\frac{1000\,{\rm GeV}}{m_{\tilde{q}}}\right)
<0.001 D.
\end{equation}
Where $D$ is a function of the ratio of gluino to squark mass and varies
between $1.6$ and $2.6$ for values of the ratio between $1/2$ and $2$.

For simplicity let us take the left and the right handed off-diagonal terms to be approximately equal,
$m^{2}_{\tilde{s}_{R}\tilde{d}_{R}} \simeq m^{2}_{\tilde{s}_{L}\tilde{d}_{L}}$. For the fine-tuning we then have,
\begin{equation}
F^{x^{2}_{\tilde{s}\tilde{d}}}_{K^{0}\leftrightarrow\bar{K}^{0}} \,\approx\,\frac{m^{2}_{\tilde{q}}}{m^{2}_{\tilde{s}\tilde{d}}}
\,\approx\,
500\left(\frac{2}{D}\right)\left(\frac{1000\,{\rm GeV}}{m_{\tilde{q}}}\right).
\label{eq:2.8}
\end{equation}
As above this fine-tuning decreases with increasing superpartner masses due to decoupling. However, the decrease is somewhat
slower (cf. the red line in Fig.~\ref{fig:finetuningindi}).

Looking at the different fine-tunings plotted in Fig.~\ref{fig:finetuningindi} we can clearly see that an increasing overall mass scale makes
the electroweak
fine-tuning worse while the ones arising from flavour (and CP) get relaxed. This suggests, that when different fine-tunings are combined the
minimal value should be in an intermediate
TeV-scale region. This is what we will look at in the next sections.

\section{Interplay between the electroweak and flavour fine-tunings}\label{sect:combine}
Having identified two (or even several) quantities that require fine-tuning we need a measure for the combined degree of unnaturalness.

The question of fine-tuning is related to the amount of parameter space available to reproduce the observed quantities (see~\cite{strumia}
for a discussion).
For a more fine-tuned parameter the parameter space producing a result of the right order of magnitude is small.
However, this parameter space can be small for two reasons 1) The measurement of the quantity in question is simply very precise, or
2) The value of the quantity is truly unnatural. For a fine-tuning we are only interested in the latter. Let us consider the case where a quantity
is unnaturally small.
Then the precision of the measurement is factored out if one asks how much of the parameter space reproduces the measured value or below.
This amount of parameter space relative to the natural volume of parameter space is measured by the typical fine-tuning measures
(see Section~\ref{sect:fl-ew}):
taking the natural parameter space interval (or volume), $1/F$ is the fraction of this interval (or volume) that gives a value of the observable with the
measured value or below. Alternatively one can say, drawing random values from the parameter space, the probability is $\sim 1/F$,
to obtain a parameter value
that gives a value for the observable equal to the measurement or below.
This probability interpretation also compels us to bound the possible fine-tuning from below by $1$. Practically this can be achieved by
taking the maximum of
the calculated fine-tuning and $1$.

This probabilistic picture naturally generalizes to a fine-tuning in several variables and observables.
If we have two independent fine-tunings, $F_{1}$ and $F_{2}$ in terms of two independent parameters, the resulting fine-tuning is given by the product,
\begin{equation}
\label{multi}
F_{tot}=F_{1}F_{2}.
\end{equation}
Obviously this can be generalized to multiple fine-tunings.

The question is, however, which parameters and observables one considers to be independent. This is basically a question of the prejudice
about the underlying fundamental model. For example in {\it ordinary} (or minimal) gauge mediation \cite{Giudice:1998bp},
all non-trivial soft mass parameters are determined by a single parameter\footnote{here ${\cal F}$ is the F-term VEV in SUSY-breaking sector and $M$ is the messenger mass.}
 ${\cal F}/M$ 
(plus $\tan\beta$)
Therefore, one can consider the fine-tuning in terms of this single parameter. In the CMSSM however, one already has several "independent"
soft mass parameters for the sfermion masses, $m_0^2$, the gaugino masses, $M_i$, and the $A$-terms, $A_0$,
and one could look at fine-tunings in all of them.

As we can see from Eq.~\eqref{multi} the
value of fine-tuning grows rapidly when we have multiple fine-tuned
observables.
Therefore assumptions
on the underlying number of independent parameters/observables crucially influence the result.
This is, however, what one should expect, because relations between different parameters is what makes a model more predictive and
potentially less fine-tuned.

Nevertheless one should take care when comparing between the numerical values in the fine-tuning in one or in several parameters.
Already a couple of fine-tunings of the order of 5, which one would not consider to be a serious fine-tuning, can increase the
numerical values of the fine-tuning parameters by more than an order of magnitude.

\begin{figure}[t!]
\begin{center}
\subfigure[]{
\includegraphics[width=0.47\textwidth]{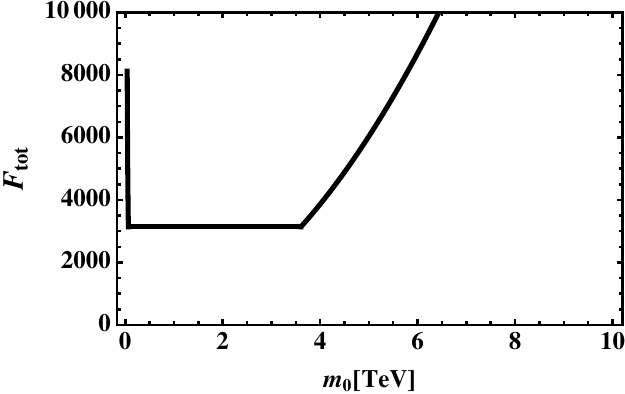}
\label{plot1}}
\hspace{0.2cm}
\subfigure[]{
\includegraphics[width=0.47\textwidth]{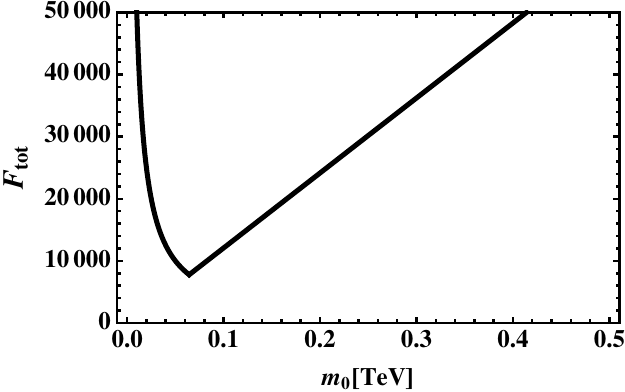}
\label{plot2}}
\\
\subfigure[]{
\includegraphics[width=0.47\textwidth]{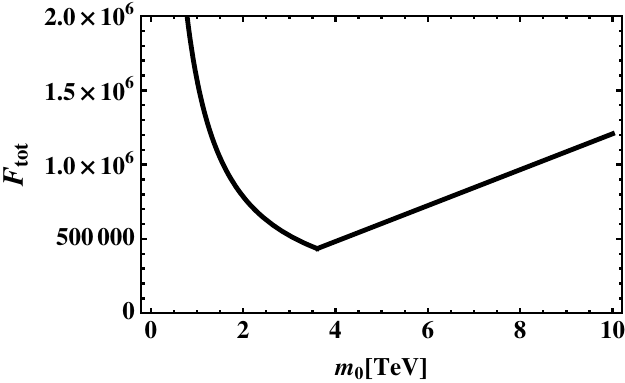}
\label{plot3}}
\end{center}
\caption{Total fine-tuning in three different scenarios as a function of the overall soft SUSY breaking scalar scale $m_{0}$. We combine the
fine-tuning for the electroweak scale depending on $m^{2}_{H_{u}}$ with the following: (a)  the flavour changing decay $\mu\to e\,\gamma$
with respect to the off-diagonal slepton mass term $m^{2}_{\tilde{\mu}_{R}\tilde{e}_{R}}$. (b)  $K^{0}\leftrightarrow\bar{K}^{0}$
oscillations with respect to the off-diagonal squark mass $m^{2}_{\tilde{s}\tilde{d}}$ (note the different scale of the axis for $m_{0}$).
(c)  both $\mu\to e\,\gamma$ and $K^{0}\leftrightarrow\bar{K}^{0}$ fine-tunings.
Note, that some care should be taken when comparing the numerical values for the resultig fine-tuning
between different approaches/models with different numbers of tuned observables and parameters.}
\label{fig:finetuning}
\end{figure}

Let us now return to our
example which combines the electroweak and flavour fine-tuning.
As discussed above, fine-tuning depends on the choice of parameters.
Here, we consider several particularly simple gravity mediation motivated choices. For all scenarios we take the masses of all
scalars to be characterised
by a scale $m_{0}$.
We now consider the combined fine-tuning $F_{tot}= F^{m^{2}_{H_{u}}}_{m^{2}_{Z}} \times F_{flavour}$.
Here the first term, $F^{m^{2}_{H_{u}}}_{m^{2}_{Z}}$, is
the fine-tuning factor necessary to achieve electroweak symmetry breaking,
characterized by the observable $m^{2}_{Z}$ with
respect to the up-type Higgs mass (in the limit of large $\tan\beta$), and the second term, $F_{flavour}$, characterises
fine-tunings for flavour observables and parameters.
In Fig.~\ref{plot1} we show the combination $F_{tot}$ with the second factor $F_{flavour}$ representing the flavour changing
decay $\mu\to e\,\gamma$ with respect to the off-diagonal slepton
mass term $m^{2}_{\tilde{\mu}_{R}\tilde{e}_{R}}$. Fig.~\ref{plot2} shows the combination $F_{tot}$ where $F_{flavour}$
 characterises $K^{0}\leftrightarrow\bar{K}^{0}$
oscillations with respect to the off-diagonal squark mass $m^{2}_{\tilde{s}\tilde{d}}$.
And finally, Fig.~\ref{plot3} is the combination
of all three fine-tunings together which are treated as independent.

Let us briefly comment on the kinks and and flat parts in the curves of Fig.~\ref{fig:finetuning} (the same behaviour occurs in our
later Fig.~\ref{fig:finetuningedm}). The kinks arise from our choice (motivated by the probabilistic argument given above) that the fine-tuning,
can never be smaller than 1.
Hence the expressions for individual measures of finetunning $F$ in Eqs.~\eqref{eq:2.4},\eqref{eq:2.6},\eqref{eq:2.8} are bounded from below by 1
in the product formula \eqref{multi}.
The flat part in Fig.~\ref{plot1} arises because the increase in fine-tuning in the Higgs mass
$\propto m^{2}_{H_{u}} \,\sim m_0^2$ in \eqref{eq:2.4}
is precisely compensated by the decrease in the flavour fine-tuning $\propto 1/m^2_{\tilde{l}_{R}} \,\sim 1/m_0^2$ in \eqref{eq:2.6}.
The compensation between these
two effects is exact in the simple approximation we have used. The flat region disappears when either finetunning effect approaches 1.

\medskip

Comparing the three different setups shown in Fig.~\ref{fig:finetuning}
it is clear that the combined fine-tuning in all cases is quite large. Moreover, we see that its overall size as well as the dependence on the parameter $m_{0}$ strongly depends on the choice of independent parameters.
Nevertheless we can see that both in Fig.~\ref{plot1} and \ref{plot3} there is a fairly broad minimum  which
occurs for scalar masses in the range of a few TeV.
Therefore from such a combined fine-tuning perspective in flavour and electroweak observables
the SUSY mass scales in the range of a few TeV are not
disfavored.

We also note that due to the weaker dependence on the squark masses the in principle stronger constraint from kaon mixing gives a less strong pull
in the fine-tuning argument. Therefore the electroweak fine-tuning dominates, resulting in a minimum at fairly low values of the soft masses.

\begin{figure}[t!]
\centerline{\includegraphics[width=0.47\textwidth]{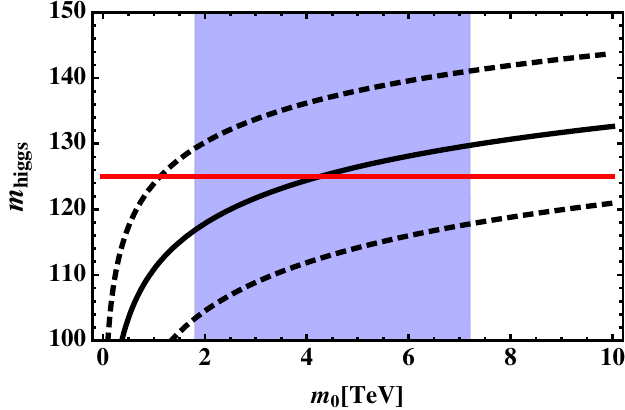}}
\caption{Higgs masses as a function of $m_{0}=m_{\tilde{t}}$ for three different values of $X_{t}=A_{t}-\mu \cot\beta$ and $\tan\beta$. From bottom to top $X_{t}=2m_{0}$ (dashed) and $X_{t}=0$ (solid) with $\tan\beta=45$ and $X_{t}=0$ with $\tan\beta=3$ (dashed).
The region where the fine-tuning is less then two times its minimal value (cf. Fig.~\ref{plot3}) is marked in light blue. The value $125\,{\rm GeV}$
is marked with a red line.}
\label{fig:higgs}
\end{figure}

Considering the combined fine-tuning measure has interesting consequences also for the
Higgs mass. In Fig.~\ref{fig:higgs} we show the expected Higgs mass (in the large $\tan\beta$ limit) in the region preferred by fine-tuning in the scenario where we have combined all the considered fine-tunings (cf. Fig.~\ref{plot3}).
In light blue we have marked the region in which the fine-tuning is less than two times its minimal value. We can see
that a $125\,{\rm GeV}$ Higgs (marked in red) is not at all that unexpected from this criterion.

\begin{figure}[t!]
\centerline{\includegraphics[width=0.47\textwidth]{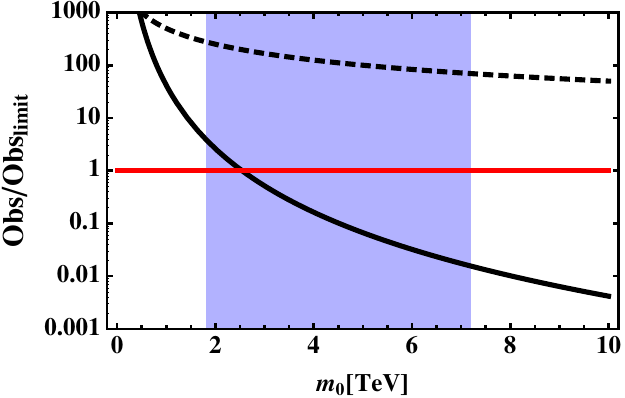}}
\caption{Expected size of the flavour changing observable compared to the current limits as a function of the overall soft scalar scale $m_{0}$. The solid line is for ${\rm BR}(\mu\to e\,\gamma)$
the dashed line for $K^{0}\leftrightarrow\bar{K}^{0}$ mixing. The region where the fine-tuning is less then two times its minimal value (cf. Fig.~\ref{plot3}) is marked in light blue. The current limit $1$ is marked with a red line.}
\label{fig:flavor}
\end{figure}

Similarly one can check for the naturally expected size of the flavour violating processes discussed above. This is shown in Fig.~\ref{fig:flavor}.
In the region where the expectation is larger than the current limit some fine-tuning of the off-diagonal elements is needed.
Fine-tuning then suggests that the value should be close to the measured upper limit in these regions. Similarly naturalness suggests that in the region
where the observed limit is not exceeded a value close to the plotted value is attained.
Overall in the whole fine-tuning preferred region we expect sizable flavour effects.

\section{CP}\label{sect:CP}

The soft SUSY breaking terms also allow for six CP violating phases. A common phase convention is to take $B\mu$ to be real.
Then the six phases can be taken as
${\rm Arg}(A^{*}_{d}M_i)$,
 with $i=1,2,3$ for the three gaugino mass parameters and
${\rm Arg}(\mu A_{\alpha})$, $\alpha=d,u,l$ for the three different $A$-terms, taken to be proportional to the Yukawa matrices.

The issue of CP violation is somewhat orthogonal to that of flavour violation. Indeed even models of gauge mediation,
which automatically have no flavour violation, can have CP violating phases associated with the combination of $\mu$ and gaugino masses,
(although these are often not considered; note, however, Ref.~\cite{feng} where this has been
suggested as a probe of gauge mediation with the heavy scalars suggested by a 125 GeV Higgs).

The CP violating SUSY phases can be sensitively tested in the (CP-violating) electric dipole moments of the electron, the neutron and the mercury nucleus.
In general there is a fairly complex dependence on the different mass and mixing parameters. Here we take a
simplified approach where we take all relevant phases to be of the same order $\phi$ and
all soft masses are again set to be of the order $m_{0}$.
Very roughly the different dipole moments are given by~\cite{Steve*2},
\begin{eqnarray}
\label{EDMs1l}
d_{e}\!\!&\sim&\!\! 10^{-25} e\,{\rm  cm}\times\sin\phi \left(\frac{300\,{\rm GeV}}{m_{0}}\right)^{2}\left(\frac{\tan\beta}{3}\right)<1.05\times 10^{-27}e\,{\rm cm}
\\\nonumber
d_{n}\!\!&\sim& \!\!10^{-24} e\,{\rm  cm}\times\sin\phi \left(\frac{300\,{\rm GeV}}{m_{0}}\right)^{2}\left(\frac{\tan\beta}{3}\right)<2.9\times 10^{-26}e\,{\rm cm}
\\\nonumber
d_{Hg}\!\!&\sim& \!\!10^{-26} e\,{\rm  cm}\times\sin\phi \left(\frac{300\,{\rm GeV}}{m_{0}}\right)^{2}\left(\frac{\tan\beta}{3}\right)<2.1\times 10^{-28}e\,{\rm cm},
\end{eqnarray}
with the current limits~\cite{current,Raidal:2008jk} given on the right hand side of the equations.

We take $\sin\phi$ to be the tuned parameter and perform the same
procedure
as before.
We then have,
\begin{equation}
\label{finetuneedm}
F^{CP}_{d}\,\sim \frac{1}{\sin\phi}\,\sim100 \left(\frac{\tan\beta}{3}\right)\,\left(\frac{300\,{\rm GeV}}{m_{0}}\right)^{2}
\,\sim 100 \,\left(\frac{300\,{\rm GeV}}{m_{0}}\right)^{2},
\end{equation}
where
for our order of magnitude estimate we have focussed on the electron EDM.

\begin{figure}[t!]
\begin{center}\subfigure[]{
\includegraphics[width=0.47\textwidth]{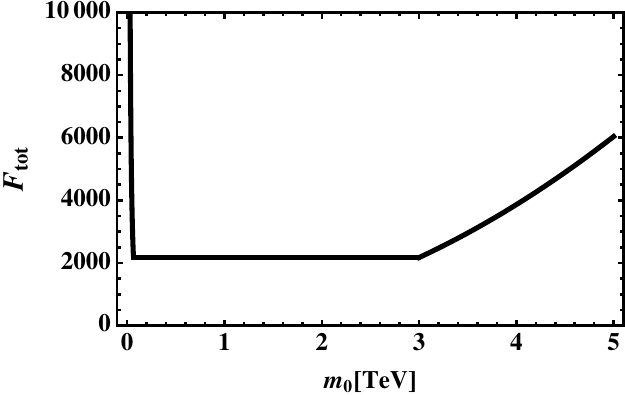}
\label{plot1edm}}
\hspace{0.2cm}
\subfigure[]{
\includegraphics[width=0.47\textwidth]{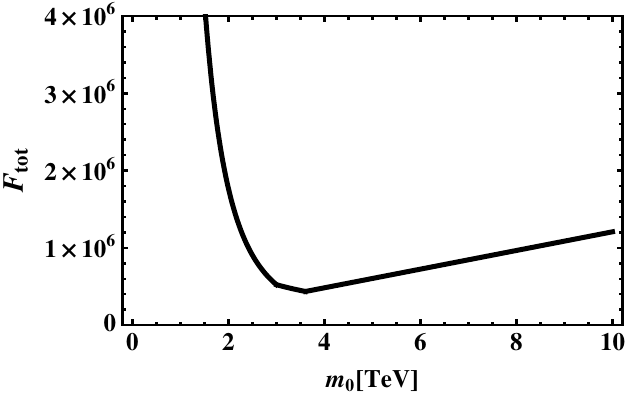}
\label{plot2edm}}
\end{center}
\caption{Total fine-tuning in two different scenarios considering CP violating quantities, as a function of the overall soft SUSY breaking scalar scale $m_{0}$. In the left panel we combine the
fine-tuning for the electroweak scale depending on $m_{H_{u}}$ with the fine-tuning of the phase $\phi$ required to agree with the electric dipole moment constraints. In the right panel (note the different scale on the axis for $m_{0}$) we combine all the finetunings discussed in this notes, electroweak, flavour ($\times 2$) and the CP violating phases. Note, that some care should be taken when comparing the numerical values for a fine-tuning
between different numbers of tuned observables and parameters.}
\label{fig:finetuningedm}
\end{figure}

The resulting fine-tuning combinations are shown in Fig.~\ref{fig:finetuningedm}. The results are fairly similar to the flavour case.
Again we find that the fine-tuning preferred region extends to a few TeV. The consequences for the Higgs and flavour quantities are therefore analogous to those in the previous section.

\begin{figure}[t!]
\centerline{\includegraphics[width=0.47\textwidth]{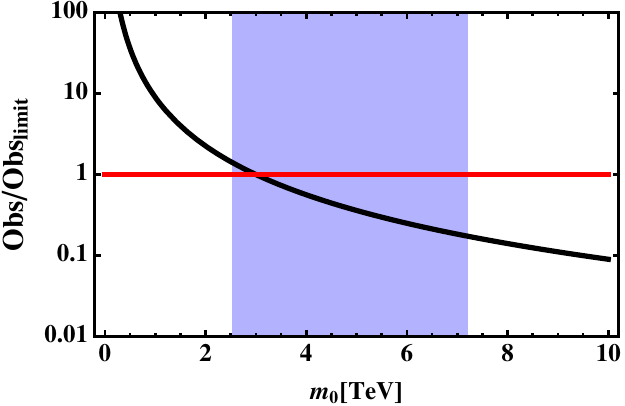}}
\caption{Expected size of the CP violating electron electric dipole moment compared to the current limits as a function of the overall soft scalar scale $m_{0}$. The region where the fine-tuning is less then two times its minimal value (cf. Fig.~\ref{plot2edm}) is marked in light blue. The current limit $1$ is marked with a red line.}
\label{fig:edm}
\end{figure}

We can also take a look at the expected size of the electric dipole moments. This is shown in Fig.~\ref{fig:edm}

Let us briefly comment on the $\tan\beta$ dependence. Our plots are for a fairly low value of $\tan\beta\sim 3$.
Increasing the value of $\tan\beta$ increases the required fine-tuning of the CP phase and therefore the overall fine-tuning.
The preferred region with minimal fine-tuning moves to the right. The shift is, however not dramatic, increasing to $\tan\beta\sim 30$
the masses increase by a factor of order two. The effect on the observable EDMs is even less severe, because larger $\tan\beta$ increases their value counteracting the decrease due to the higher masses.

Our analysis of CP violation above was based on bounds in Eqs~\eqref{EDMs1l} arising from one-loop diagrams involving scalars.
At large values of scalar masses $m_0$, these one-loop EDMs are suppressed. One can ask what would be the effect of two-loop contributions
to EDMs, the most important of which would be the contribution  to the neutron EDM of the Weinberg dimension-6 purely gluonic operator
\cite{Chang:1998uc,Steve*2}. This effective operator is generated due to stops circulating in the loop and their effect on
the EDM can become
significant relative to the one-loop contribution if the stops are much lighter than the squarks of the first two generations.
However for the uncomplicated generic SUSY models considered here we take all scalar masses to be of similar size, and there is no enhancement
of two-loop effects compared to the leading-order contributions in \eqref{EDMs1l}. Quite generally, as already mentioned in the introduction,
our aim is to avoid specialising to
particular SUSY models with specific in-built hierarchies of scales or to corners of their parameter
spaces.

So far we have also only considered a simple situation with one overall parameter that controls the magnitude of all CP violating SUSY phases.
In principle one could imagine some or all of the possible CP phases are independent.
Then one would need to consider the combined fine-tuning in all of those parameters (taking into account the constraints on different EDMs).
As EDMs generally prefer larger SUSY masses this pushes the fine-tuning preferred region upwards. However, the shift is typically not dramatic.

\section{Additional fine-tunings and fine-tuned parameters}
So far we have focussed on the fine-tuning arising from electroweak symmetry breaking as well as the strong constraints on flavour
and CP observables.
In general one could include further observables in the analysis, e.g. $(g-2)_{\mu}$ and dark matter.

\medskip

For $(g-2)_{\mu}$ one could take two different viewpoints. One is to require that the SUSY contribution to $(g-2)_{\mu}$ does not exceed
the limits set by experiment. At very low masses contributions to $(g-2)_{\mu}$ are typically too large requiring fine-tuning to make it small.
This makes $(g-2)_{\mu}$ an additional variable that prefers a larger SUSY breaking scale. However, direct constraints have already pushed
us into a regime where this is not a dramatic effect anymore. Alternatively one could require that the discrepancy between the currently
observed value and the Standard Model prediction is explained by SUSY. In this case a too large value of the generic SUSY scale is disfavoured,
as one would need to tune certain masses, e.g. smuon and bino  masses, to be small in order to be achieve a sufficiently large SUSY contribution
to $(g-2)_{\mu}$.
Pending a more detailed analysis, we do not expect a qualitative change in the conclusions obtained in the previous sections
(details will also depend on the approach taken).

\medskip

We now briefly comment on dark matter\footnote{The finetuning originating from dark matter was first considered in~\cite{Ellis:2001zk}.}.
In SUSY, thermally produced neutralino dark matter is generically produced with the right abundance when $m_{\chi^{0}}\sim 100\,{\rm GeV}$.
With increasing $m_{\chi^{0}}$, the total abundance naively increases as $\sim m^{2}_{\chi^{0}}$. Therefore high SUSY scales require an additional
mechanism to reduce this abundance. One example is co-annihilation which occurs when the neutralino mass is very close to a sfermion mass,
often the mass of the stau. This can be viewed as an additional fine-tuning in the parameter $m_{\tilde{\tau}}-m_{\chi^{0}}$ which is required
to be small compared to the overall SUSY scale. This leads to a preference for a low overall SUSY scale.
Including this will shift the fine-tuning-preferred mass scale somewhat downwards.
Let us note, however, that this is also an example where one can clearly see the crucial dependence on what one considers fundamental parameters
and in particular at which scale they are evaluated. Indeed, in the CMSSM the fine-tuning in the co-annihilation region is smaller
than expected, because there $m_{0}\ll m_{1/2}$ and therefore the stau mass is dominated by the RG running contribution from the bino and therefore
less sensitive to $m_{0}$~\cite{King}.
Parenthetically we also note that one could choose not to explain the dark matter in terms of SUSY WIMPs\footnote{Overabundances could, for example, be removed by a cosmological dilution mechanism.}, after all this is not a `must'
unlike in the case of suppressing flavour and CP violation, and in this case there is no additional finetuning.

\medskip
We note that both $(g-2)_{\mu}$ and the dark matter abundance are observables that are directly influenced by the gaugino masses. This
is in contrast to electroweak symmetry breaking and the flavour observables which most directly depend on the sfermion masses.
Therefore these variables could be used to determine an optimal value for the gaugino mass scale from fine-tuning considerations.

\medskip
In this paper we have considered one parameter to be fine-tuned for each observable. Depending on the model observables can be influenced by more
than one parameter. In this case one can consider the maximum of the fine tunings in the different directions (or the root of the sum of squares).

\section{Conclusions}
The new constraints arising from both direct SUSY searches and the hints for a Higgs at $125\,{\rm GeV}$
force the SUSY spectra to become heavier.
This effect is particularly severe for simple models such as gauge mediation
which were constructed to automatically solve the flavour problem.
With higher masses the fine-tuning
required to achieve electroweak symmetry breaking increases. At the same time a heavier spectrum of SUSY particles may ameliorate problems arising from other sources such as flavour violation and CP. In light of this we have investigated models with anarchic flavour structure and CP violation.
These models face a combination of fine-tunings from electroweak symmetry breaking, flavour and CP violation.
Interestingly in these models the combined fine-tuning gives a preference for masses in the region of a few TeV.
This is consistent with direct searches and the hints for a $125\,{\rm GeV}$ Higgs.
Moreover it suggests that observable flavour and CP violating effects may not be too far away
and direct searches may yet have a chance for discovery.

\bigskip

\section*{Acknowledgements}
We would like to thank Steven Abel, Guido Altarelli, Matthew Dolan, Michael Spannowsky and Chris Wymant for stimulating discussions and comments.
VVK gratefully acknowledges the support of the Wolfson Foundation and
 the Royal Society.

\end{document}